\documentclass[12pt,preprint]{aastex}

\shorttitle{Optical Spectroscopy of Large KBOs}
\shortauthors{Tegler et al.}

\begin{document}


\title{Optical Spectroscopy of the Large Kuiper Belt Objects \\
    136472 (2005 FY9) and 136108 (2003 EL61)}


\author{S. C. Tegler\altaffilmark{1}}
\affil{Department of Physics and Astronomy, Northern Arizona University,
    Flagstaff, AZ, 86011}
\email{Stephen.Tegler@nau.edu}

\author{W. M. Grundy}
\affil{Lowell Observatory, 1400 W. Mars Hill Road, Flagstaff, AZ, 86001}

\author{W. Romanishin\altaffilmark{1}} 
\affil{Department of Physics
and Astronomy, University of Oklahoma, Norman, OK, 73019}

\author{G. J. Consolmagno\altaffilmark{1,2}}
\affil{Vatican Observatory, Specola, Vaticana, V-00120, Vatican City State}
\affil{Department of Physics, Fordham University, Bronx, NY, 10458}

\author{K. Mogren}
\affil{Department of Physics and Astronomy, Northern Arizona University, 
Flagstaff, AZ, 86011}

\bigskip

\and

\author{F. Vilas \altaffilmark{1}} \affil{MMT Observatory, PO Box 210065, University of
Arizona, Tucson, AZ, 85721}


\altaffiltext{1}{Observer at the MMT Observatory. Observations reported here were obtained at the MMT
Observatory, a joint facility of the University of Arizona and the Smithsonian Institution.}
\altaffiltext{2}{2006-2007 Loyola Chair, Fordham University}


\begin{abstract}
We present high signal precision optical reflectance spectra of the
large Kuiper belt objects 2005 FY9 and 2003 EL61.  The spectrum of
2005 FY9 exhibits strong CH$_4$-ice bands.  A comparison between the
spectrum and a Hapke model indicates the CH$_4$ bands are shifted 3.25
$\pm$ 2.25 \AA\ relative to pure CH$_4$-ice, suggesting the presence
of another ice component on the surface of 2005 FY9, possibly
N$_2$-ice, CO-ice, or Ar. The spectrum of 2003 EL61 is remarkably
featureless.  There is a hint of an O$_2$-ice band at 5773 \AA;
however, this feature needs to be confirmed by future spectroscopic
observations of 2003 EL61 with a higher continuum signal precision,
sufficient to detect a second weaker O$_2$-ice band at 6275 \AA.
\end{abstract}





\section{Introduction}

Measuring the surface composition of Kuiper belt objects (KBOs) may
provide clues about the composition and environment of the primordial
solar nebula as well as the important evolutionary processes occurring
in the outer Solar System over the last 4.5 Gyr.  Unfortunately, there
are only a handful of KBOs that are known to exhibit ice absorption
bands in their spectra. H$_2$O-ice bands are seen in spectra of 1996
TO66 \citep{brn99}, Varuna \citep{lic01}, Quaoar \citep{jl04}, and
Orcus \citep{for04a}.  CH$_4$-ice bands are seen in spectra of Pluto
\citep{crk76,fink80,gf96}, Neptune's satellite Triton, which may be a
captured KBO \citep{crk93}, Eris \citep{brn05}, and 2005 FY9
\citep{lic06a} . There are perhaps a dozen objects that exhibit
spectra with no ice absorption bands \citep{dor03,for04b}

The recent discovery of extraordinarily bright and large KBOs opens up
a new opportunity for the physical and chemical studies of KBO
surfaces with high signal precision optical reflectance spectroscopy.
Bright objects make it possible to either carry out in-depth physical
modeling of ice and mineralogical absorption bands or set stringent
upper limits on the presence of such bands.

Here we present high signal precision optical reflectance spectra of
KBOs 2005 FY9 and 2003 EL61. Both objects are candidates for
membership in the newly defined class ``dwarf planet.''  The diameter
of 2005 FY9 is estimated at 1600 km \citep{brn06} while 2003 EL61
appears to be a highly elongated ellipsoid with axes of 1950 $\times$
2500 km \citep{rab06}.  For comparison, Pluto has a diameter of 2350
km. 2005 FY9 and 2003 EL61 are among ``scattered-near'' objects
(i.e. non-resonant, non-planet-crossing objects with Tisserand
parameters less than 3, relative to Neptune) in the Deep Ecliptic
Survey classification system \citep{ell05}.  2005 FY9 has a perihelion
distance, q, of 38.6 AU, an aphelion distance, Q, of 52 AU, a
semi-major axis, a, of 45.3 AU, an inclination angle to the ecliptic,
i, of 29$^{\circ}$, and an eccentricity of 0.15. 2003 EL61 has q $=$
35.1 AU, Q $=$ 51 AU, a $=$ 43.1 AU, i $=$ 28$^{\circ}$, and e $=$
0.18 For comparison, Pluto has q $=$ 29.7 AU, Q $=$ 49 AU, a $=$ 39.5
AU, i $=$ 17$^{\circ}$, and e $=$ 0.25.  At the time of observation,
2005 FY9 and 2003 EL61 were both near aphelion.


\section{Observations}

We obtained spectra of 2005 FY9 and 2003 EL61 on 2006 March 5 UT with
the 6.5-meter MMT telescope on Mt. Hopkins, AZ, the Red Channel
Spectrograph, and a 1200 $\times$ 800 CCD.  A 150 g mm$^{-1}$ grating
and a 1 arc sec slit width provided wavelength coverage of 5000
to 9500 \AA\  in first order, a dispersion of 6.38 \AA\ pixel$^{-1}$,
and a fwhm resolution of 20.0 \AA. A LP495 blocking filter eliminated
contamination from the second order.

There were high, thin cirrus clouds through most of the night and the
seeing was $\sim$ 0.8 arc sec. The KBOs were placed at the center of
the slit and tracked at KBO rates. 

We used the Image Reduction and Analysis Facility (IRAF) and standard
procedures \citep{mas92} to calibrate and extract one-dimensional
spectra from the two-dimensional spectral images. Specifically, the
electronic bias of each image was removed by subtracting its overscan
as well as a bias picture. Pixel to pixel sensitivity variations were
removed from each image by dividing by a normalized twilight flatfield
image. Extraction of one-dimensional spectra from the images was done
with the apall task in IRAF.  HeNeAr spectra were used to correct for
flexure and obtain an accurate wavelength calibration. Our wavelengths
are accurate to $\sim$ one-tenth of a pixel or $\sim$ 0.7 \AA. We
removed telluric bands and Fraunhofer lines from the KBO spectra by
observing the solar analog HD 112257 \citep{har82} at airmasses very
close to 2005 FY9 and 2003 EL61, and then then dividing the KBO
spectra by the normalized solar analog spectra. Typically, the airmass
difference between the KBOs and the solar analog was $\le$ 0.05.  The
2005 FY9 and 2003 EL61 spectra in the analysis below are an average of
four 10-minute and five 10-minute exposures, respectively.

\section{Results}

\subsection{2005 FY9}

We plot our reflectance spectrum of 2005 FY9 in Figure 1 (bottom).
The tick marks correspond to previously reported absorption maxima of
CH$_4$-ice (Table 1).  The middle and top spectra in Figure 1 are of
2005 FY9 \citep{lic06a} and Pluto \citep{gf96}. All spectra are
normalized to 1 at 6500 \AA. For the purpose of comparison, the
\citet{lic06a} 2005 FY9 and Pluto spectra are offset by 0.4 and 0.8,
respectively. Clearly, 2005 FY9 exhibits much deeper CH$_4$-ice bands
than Pluto.

Our spectrum differs from the \citet{lic06a} spectrum in two
ways. First, our spectrum exhibits additional subtle absorption bands
at 5400, 5800, and 6000 \AA. These bands have not been seen before in
astronomical or laboratory spectra of CH$_4$-ice.  Because the bands
are close to the wavelengths of gas phase CH$_4$ bands and because
there are strong CH$_4$-ice bands at longer wavelengths in the
spectrum of 2005 FY9, we think the new bands are due to CH$_4$-ice. In
addition, it appears the band at 6200 \AA\ is CH$_4$-ice as well. It
is in the \citet{lic06a} spectrum, but they do not identify it as
CH$_4$-ice. Second, we find a continuum slope between 5500 and 6500
\AA\ of 10\%/1000\AA\ while \citet{lic06a} find a slope of
13\%/1000\AA, with both spectra normalized to 1 at 6500 \AA.  We used
HD 112257, a G2V star, for our solar analog star whereas
\citet{lic06a} used BS 4486, a G0V star, for their solar analog star.
An examination of the V$-$R colors for G0V and G2V stars suggests the
slope difference is due to the use of different solar analog stars.

In order to place constraints on the CH$_4$ grain size as well as the
presence of any additional ice components on the surface of 2005 FY9,
we calculated model CH$_4$-ice spectra. Our models use laboratory
optical constants for pure CH$_4$-ice at 30 K \citep{grun02} and
arbitrary optical constants that absorb more at blue than red
wavelengths, and thereby reproduce the observed reddish slope for
the 2005 FY9 continuum. The ice and reddening agent were mixed in the
model at the molecular level (i.e. by means of a weighted average of
the ice and reddening agent optical constants).  We used Hapke theory
\citep{hap81,hap93} to transform the optical constants into reflectance
spectra.  We used Hapke parameters h $= $0.1, B$_o$ $=$ 0.8,
$\overline{\theta}$ $=$ 30$^{\circ}$, and P(g) $=$ a two component
Henyey-Greenstein function with 80\% in the forward scattering lobe
and 20 \% in the back scattering lobe and both lobes having asymmetry
parameter 0.63.  These values are comparable to numbers used
previously to model the surface of Pluto \citep{gb01}.

Figure 2 compares the spectrum of 2005 FY9 (black line) and a Hapke
model with a grain size of 0.55 cm (red line). Although the model fits
the core of the 8897 \AA\ band, it has too little absorption at all
the other bands. Next, we fit the core of the weaker 7296 \AA\ band
with a 1.8 cm grain size. Then, we fit the core of the even weaker
8000 \AA\ complex with a 2.5 cm grain size.  These models illustrate a
trend, weaker bands (which see deeper into the surface) aways seem to
need bigger grain sizes, or equivalently, greater optical path lengths
in CH$_4$-ice.

From the above discussion, it is clear a single grain size will not
fit the observations. In Figure 3, we present a Hapke model that uses
two grain sizes, 6 cm (97 \% by volume) and 0.1 cm (0.03 \% by
volume), in an intimate mixture.  The inability of the two-grain size
model to fit the CH$_4$-ice band at wavelengths less than 7000 \AA\ in
Figure 3 is due to the lack of laboratory optical constants.  We note
that our two-grain size model is not a unique solution. Although this
model fits the data, other models (such as courser grains underneath a
coating of finer grains) would probably fit the data just as well.  In
addition, it is important to realize that grain sizes larger than 1 cm
are probably a measure of the spacing of fractures or voids in the
solid surface rather than indicating a surface covered by golf-ball
sized particles.

Although we can fit the shape and depth of the absorptions, there is a
small but significant difference between our models and the telescope
data.  Specifically, the maxima of the CH$_4$ absorption bands in the
spectrum of 2005 FY9 reside blueward of the maxima in the pure CH$_4$
Hapke models.  In Figure 4, we present a small portion of Figure 3
centered on the 7296 \AA\ band to illustrate the shift.  Such a shift
is important because laboratory experiments show that the 8897, 8968,
and 9019 \AA\ bands in spectra of pure CH$_4$-ice are shifted blueward
by 27 \AA\ in a CH$_4$/N$_2$ $=$ 0.8 \% mixture \citep{qs97}. Another
experiment finds the 8897 \AA\ band shifts $\sim$ 17 \AA\ in a
CH$_4$/N$_2$ $=$ 20 \% ice mixture \citep{gf93}.

A cross correlation experiment provides a way to quantify the apparent
shift in Figure 3. Specifically, we wrote a Fortran program to shift
the model spectrum from $-$ 25 to $+$25 \AA\ in 1 \AA\ steps.  For
each shift, the program finds the difference between the data and
model, i.e. it calculates $(y_{d,i} - y_{m,i})$ where $y_{d,i}$ and
$y_{m,i}$ represent the ordinate values of the data and the model
spectra at wavelength i. Then, the program sums the squares of the
difference over all N wavelength points between 7000 and 9300 \AA.  In
other words, we calculate

\begin{equation}
\Delta_{shift} = \sum_{i}^{N} (y_{d,i} - y_{m,i})^2.
\end{equation}

\noindent In Figure 5, we present a plot of the sums of the squares of
the differences, i.e. $\Delta_{shift}$, as function of shift. We find
a well defined minimum at 3.25 \AA.

What is the uncertainty in the 3.25 \AA\ measurement? We note that
HeNeAr spectra enable us to calibrate the wavelengths in the 2005 FY9
spectrum to an uncertainty of $\sim$ 1/10 of a pixel or $\sim$ 0.7
\AA. Specifically, we find the average difference between the
centroids of airglow lines in our data and the corresponding
laboratory values is 0.7 \AA. A larger source of uncertainty comes
from the noise in the spectra and the broadness of the cross
correlation minimum in Figure 5.  Propagation of the noise in the data
and model spectra for the 3 \AA\ shift through equation (1) gives an
ordinate error bar for the minimum point (3.25, 2.218) in Figure 5 of
0.042.  As a result, the uncertainty in our shift measurement is
defined by $\Delta_{shift}$ $<$ 2.218 $+$ 0.042 $=$ 2.260 (i.e. the
section of the curve below the dashed line in Figure 5), corresponding
to shifts between 1 and 5.5 \AA. In short, we find the CH$_4$-ice
bands in the spectrum of 2005 FY9 are blueshifted relative to the
model Hapke spectrum by 3.25 $\pm$ 2.25 \AA.

\citet{lic06a} gave blueshifts of 2 - 6 \AA\ for CH$_4$ bands in their
Table 1, but they questioned the reality of the shifts because they
are so small.  Our independent measurement of shifts with similar
magnitude bolsters the case for the reality of the small shifts.  In
fact, when we take the \citet{lic06a} data and do a cross correlation
experiment between our spectrum and their spectrum, we find no
measurable shift between the spectra; but when we perform a cross
correlation experiment between their spectrum and the Hapke model, we
find their CH$_4$-ice bands are blue shifted 2 \AA\ relative to the
model.  Our spectrum and the \citet{lic06a} spectrum are consistent.

\subsection{2003 EL61}

We plot our reflectance spectrum of 2003 EL61 in Figure 6.  The
continuum has a slope of essentially zero, consistent with B$-$V $=$
0.63 $\pm$ 0.03 and V$-$R $=$ 0.34 $\pm$ 0.02
\citep{rab06}. Evidently, tholins are not present on the surface of
2003 EL61 at an abundance seen on 2005 FY9. (Recall that both objects
were observed on the same night and compared against the same solar
analog spectrum.)  We find no evidence of CH$_4$-ice bands, but we can
set an upper limit on the thickness of a global glaze of CH$_4$-ice by
applying Beer's Law,

\begin{equation}
{\left(I \over I_o\right)}_{\lambda} = e^{-2a_{\lambda}t},
\end{equation}

\noindent to the continuum near the 8897 \AA\ band. In the above
equation, t is the thickness of the glaze, a$_{\lambda}$ are the
Lambert absorption coefficients for CH$_4$-ice \citep{grun02}, the
factor of two comes about because light passes through the glaze once
on the way in and once on the way back out, (I/I$_o$)$_{\lambda}$ is
the ratio of light leaving the glaze to the light incident on the
glaze. For (I/I$_o$)$_{\lambda=8897}$ $=$ 0.97, we find t $<$ 0.3 mm.

H$_2$O-ice bands are seen in near-infrared spectra of 2003 EL61
\citep{bar06}. Therefore, it is reasonable to look for evidence of
solid-state photolytic or radiolytic chemistry
\citep{jon90,jon97,del97} by seeing if O$_2$-ice is present on the
surface of 2003 EL61. In Figure 6, we superimpose the 5773 and 6275
\AA\ O$_2$-ice bands in spectra of Ganymede \citep{cal97} on top of
our 2003 EL61 spectrum. There is a tantalizing dip at 5773 \AA\ in the
spectrum of 2003 EL61, but no dip at the weaker 6275 \AA\ band.  The
continuum signal precision at the position of these bands is
about 150 and so if the 5773 \AA\ band is real, it would take a
continuum signal precision of about 300 to detect the 6275 \AA\
band.

We estimate an upper limit to the thickness of a global glaze of
O$_2$-ice by assuming the possible feature at 5773 \AA\ is real and
then applying Beer's Law as we did for CH$_4$-ice.  For
(I/I$_o$)$_{\lambda=5573}$ $=$ 0.99 and a$_{\lambda=5773}$ $=$ 5.5
cm$^{-1}$, we find t $<$ 0.01 mm. Our value for a$_{\lambda=5773}$
comes from an integrated absorption coefficient of 1.1 $\times$ 10$^3$
cm$^{-2}$ and a band fwhm of $\sim$ 200 cm$^{-1}$ \citep{lan62}.


\section{Conclusions}

Our spectrum of 2005 FY9 exhibits strong CH$_4$-ice bands in agreement
with the spectrum of \citet{lic06a}. From a comparison of our spectrum
and a Hapke model, we find the 2005 FY9 CH$_4$-ice bands are shifted
3.25 $\pm$ 2.25 \AA\ blueward relative to the positions of pure
CH$_4$-ice bands. The shift could be due to the presence of another
ice-component, possibly N$_2$-ice, CO-ice, or Ar.  Future higher
resolution spectra of the individual CH$_4$ bands should determine if
the weaker bands which penetrate deeper into the surface exhibit
different shifts than the stronger bands which do not penetrate as deep
into the surface. Such observations could provide a technique to
measure the CH$_4$ concentration relative to the other ice component
as a function of depth below the surface of 2005 FY9. In addition,
it is highly desirable to obtain a measure of the shift as a function
of rotational longitude on 2005 FY9 for a particular CH$_4$-ice
band. Such a technique could provide a measure of inhomogeneity on the
surface of 2005 FY9. \citet{gf96} made such measurements for the 8897
\AA\ band on Pluto and found that the blueshift varied from 0 to 10
\AA\ over the surface.

On Pluto, it appears N$_2$ is sufficiently mobile to form textures
that allow incoming photons to travel several centimeters through the
very transparent, polycrystalline N$_2$-ice before the photons are
scattered off inclusions or grain boundaries. The long pathlengths
make it possible for a small amount of CH$_4$ relative to N$_2$ to
give deep CH$_4$ absorption bands. The mobility of CO is only a little
less than N$_2$, and the mobility of Ar is only a little less than
CO. Therefore, it is possible that CO-ice or Ar-ice could provide the
host matrix for long optical pathlengths. It is also possible the host
matrix is a combination of N$_2$, CO, and Ar. A suite of laboratory
experiments that measure the shifts of CH$_4$ absorption maxima for
different concentrations of CH$_4$ relative to N$_2$, CO, and Ar would
be highly valuable for the future interpretation of observational
data.

Our spectrum of 2003 EL61 is remarkably featureless. There is a hint
of a band at 5773 \AA\ possibly due to O$_2$-ice.  Future work will
attempt to double the signal precision of the continuum to at
least 300, which is the level that appears to be required for any
possible detection of the weaker O$_2$-ice band at 6275 \AA.  The
detection of O$_2$-ice in significant quantities on KBOs could be of
considerable interest in the distant future as a source of spacecraft
fuel.
\bigskip

\centerline{\bf Acknowledgments}

SCT, WR, and KM gratefully acknowledge support from NASA Planetary
Astronomy grant NNG06G138G to Northern Arizona University and the
University of Oklahoma. WMG gratefully acknowledges support from NASA
Planetary Geology and Geophysics grant NNG04 G172G to Lowell
Observatory. GJC gratefully acknowledges support from Fordham
University. We thank Steward Observatory for the allocation of
telescope time on the MMT. We thank Dr. J. Licandro for providing us
with his spectrum of 2005 FY9. We thank an anonymous referee for a
careful review of our manuscript.

\clearpage

\begin{figure}
\includegraphics[angle=0,scale=.80]{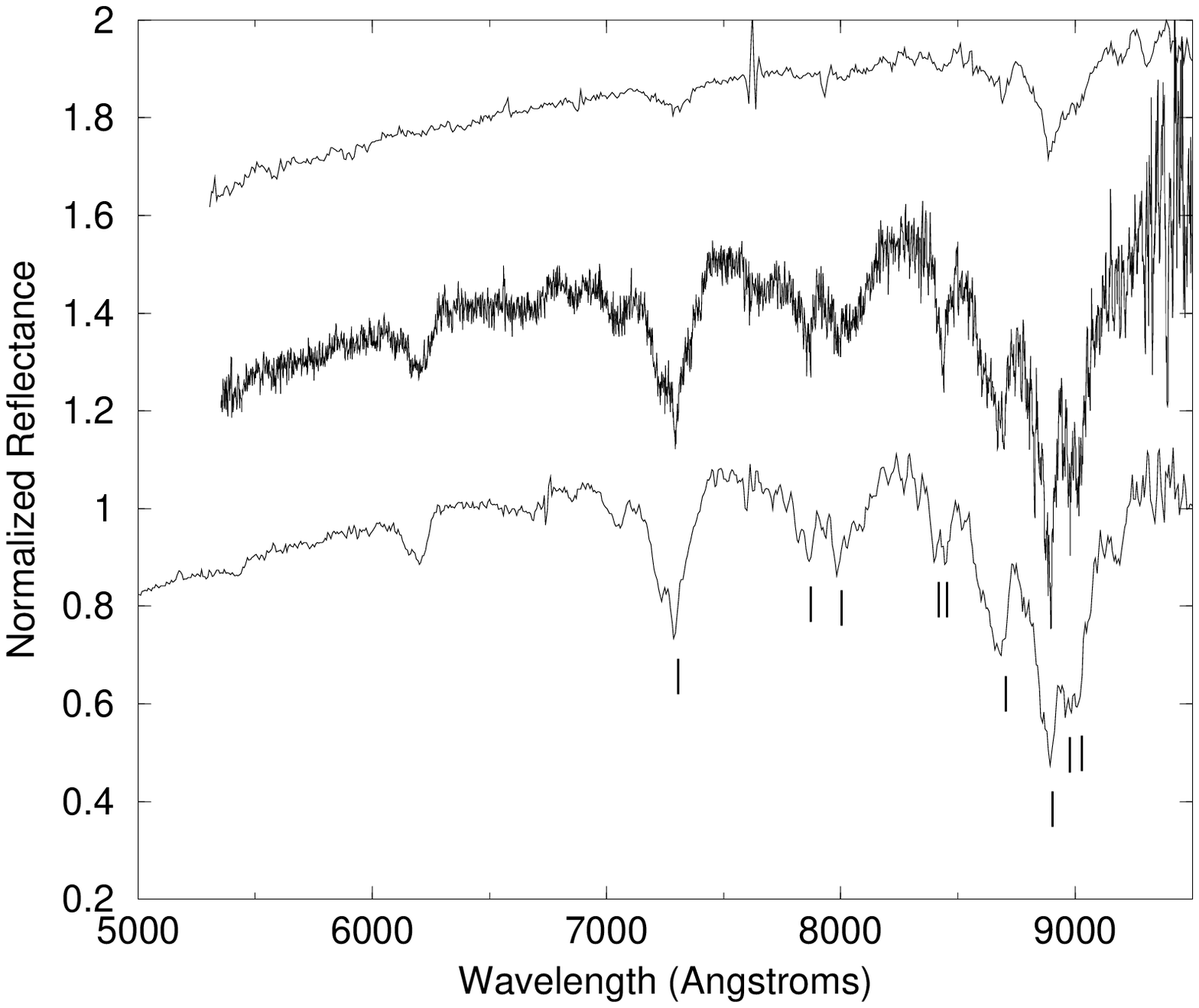}
\caption{Our normalized reflectance spectrum of 2005 FY9 (bottom).
Tick marks indicate the position of previously reported CH$_4$-ice
bands, see Table 1.  The middle spectrum is also of 2005 FY9
\citep{lic06a}. The top spectrum is of Pluto \citep{gf96}. All spectra
are normalized to 1 at 6500 \AA.  The \citet{lic06a} and Pluto spectra
are offset by 0.4 and 0.8, respectively. Clearly, 2005 FY9 has deeper
CH$_4$-ice bands than Pluto. Previously unreported CH$_4$-ice bands
appear near 5400, 5800, and 6200 \AA\ in our spectrum of 2005 FY9.}
\end{figure}

\begin{figure}
\includegraphics[angle=0,scale=.80]{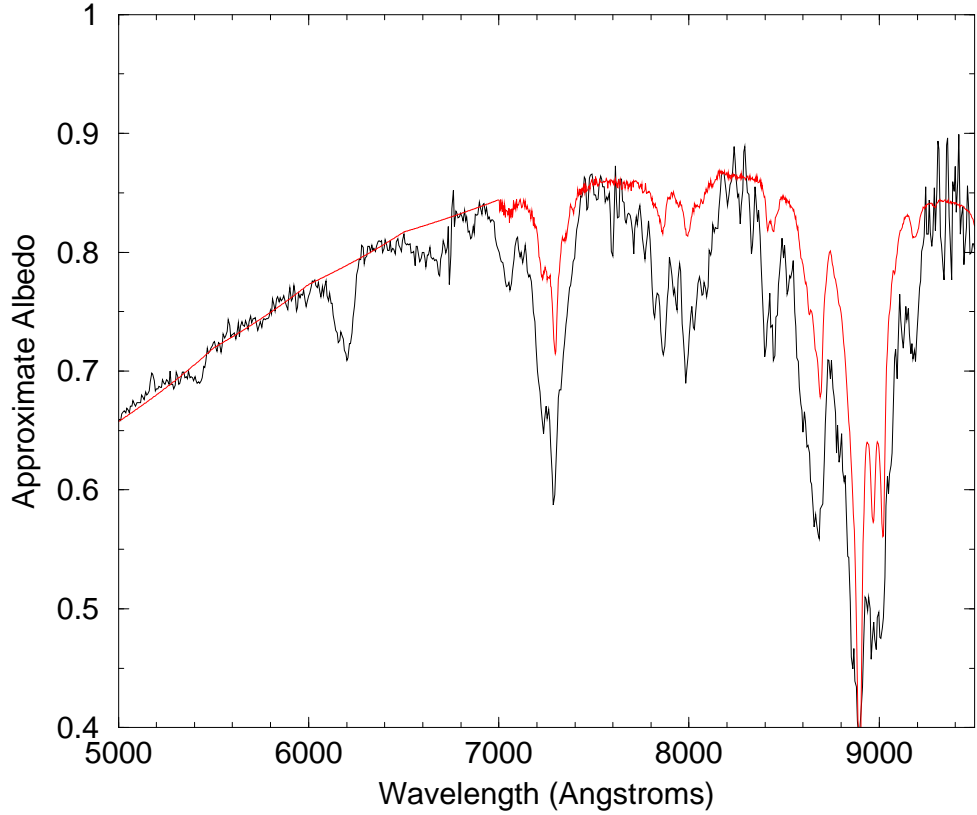}
\caption{Approximate albedo spectrum of 2005 FY9 (black line) and a
Hapke model with a reddening agent and a CH$_4$-ice grain size of 0.55
cm (red line). We assume an albedo at 6500 \AA\ of 0.8. The model fits
the core of the 8897 \AA\ band, but has too little absorption at all
other bands. There are no laboratory absorption coefficients to fit
CH$_4$ bands with $\lambda$ $<$ 7000 \AA.}
\end{figure}

\begin{figure}
\includegraphics[angle=0,scale=.80]{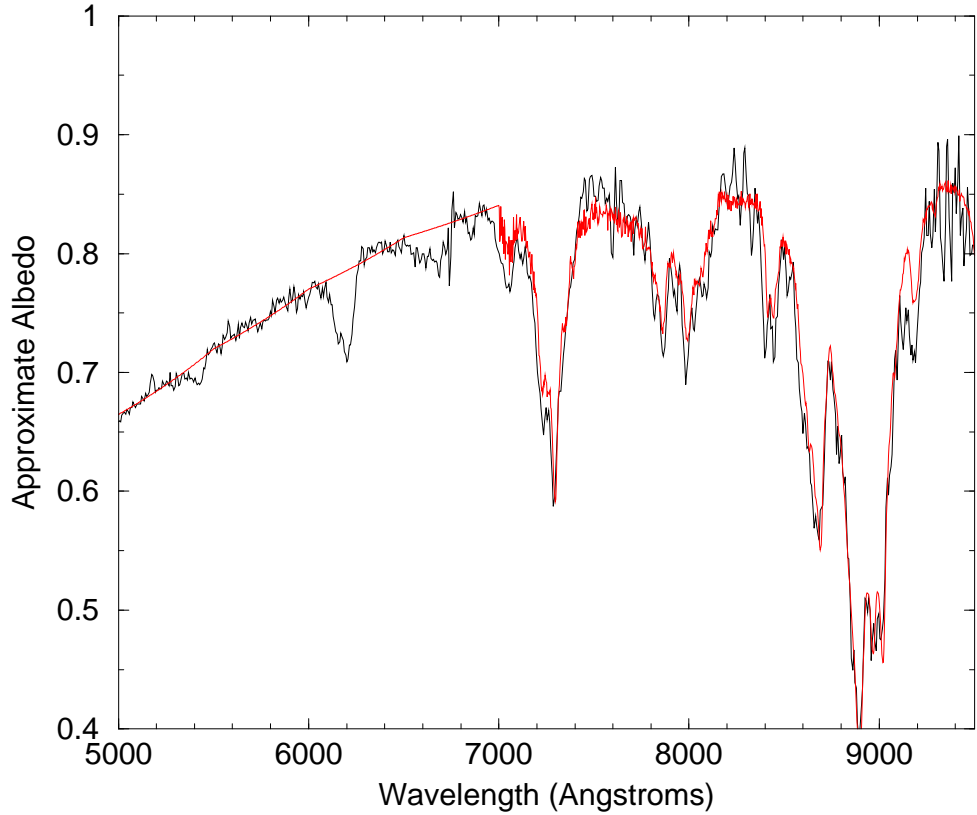}
\caption{Approximate albedo spectrum of 2005 FY9 (black line) and a
Hapke model with a reddening agent and two CH$_4$-ice grain sizes, 6
cm and 0.1 cm (red line). We assume an albedo at 6500 \AA\ of
0.8. Grains sizes larger than a cm probably provide a measure of the
spacing between fractures or voids in the surface rather than
indicating a surface covered with golf-ball sized particles.  There
are no laboratory absorption coefficients to fit CH$_4$ bands with
$\lambda$ $<$ 7000 \AA}
\end{figure}

\begin{figure}
\includegraphics[angle=0,scale=.80]{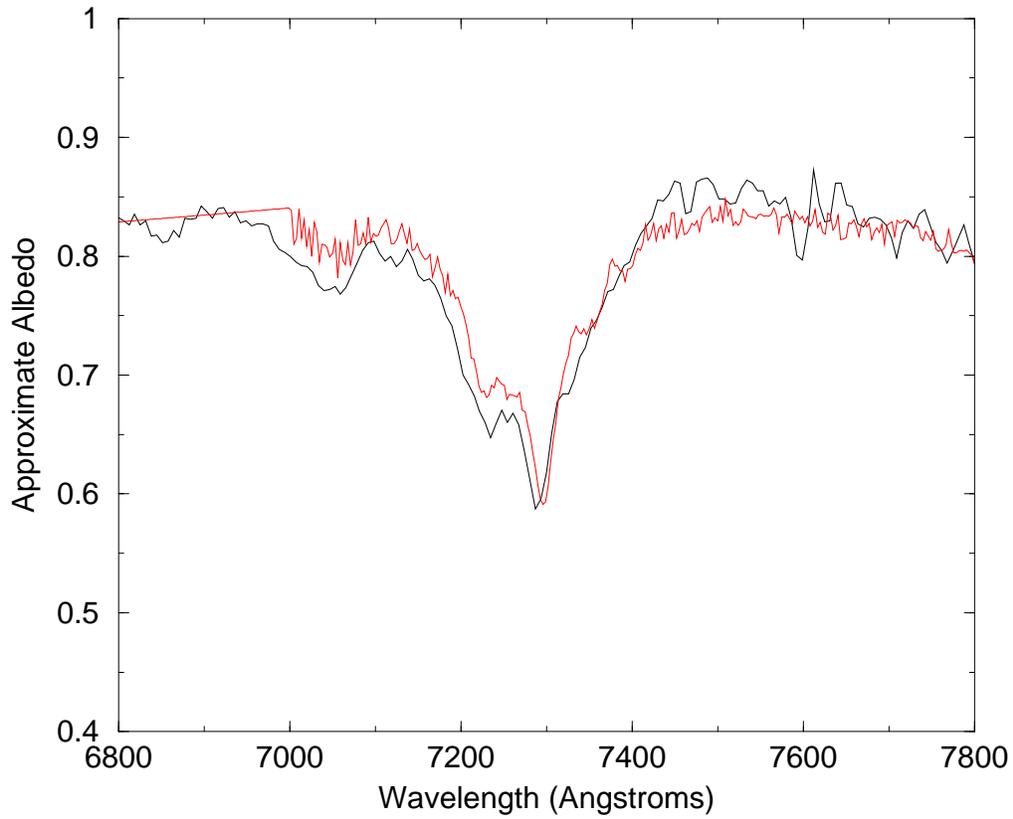}
\caption{The minimum of the 7296 \AA\ CH$_4$-ice band in 2005 FY9
(black line) is blueshifted relative to the minimum of the pure
CH$_4$-ice Hapke model (red line). The other bands in 2005 FY9 show a
similar shift. The shift may be due to the presence of another ice
component on the surface of 2005 FY9.}
\end{figure}

\clearpage

\begin{figure}
\includegraphics[angle=0,scale=.80]{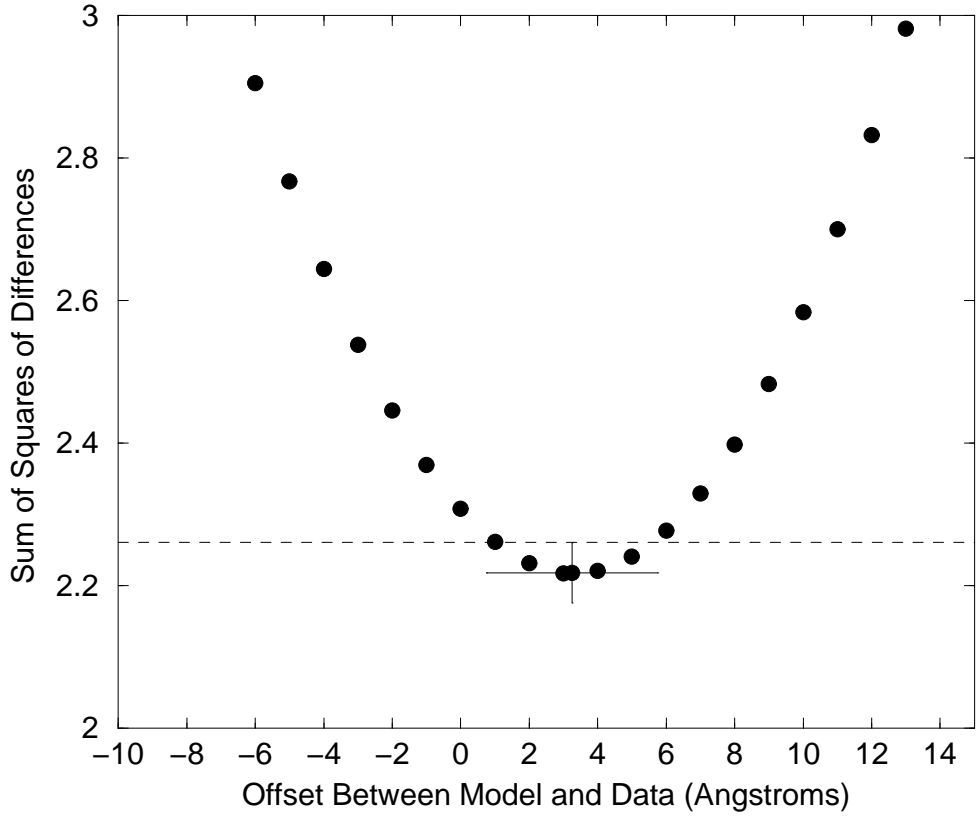}
\caption{Result of cross-correlation experiment between the spectra of
2005 FY9 and the pure CH$_4$-ice Hapke model in Figure 3. The
difference between the two spectra reach a minimum at a 3.25 $\pm$
2.25 \AA\ blueshift of the data relative to the model.}
\end{figure}

\begin{figure}
\includegraphics[angle=0,scale=.80]{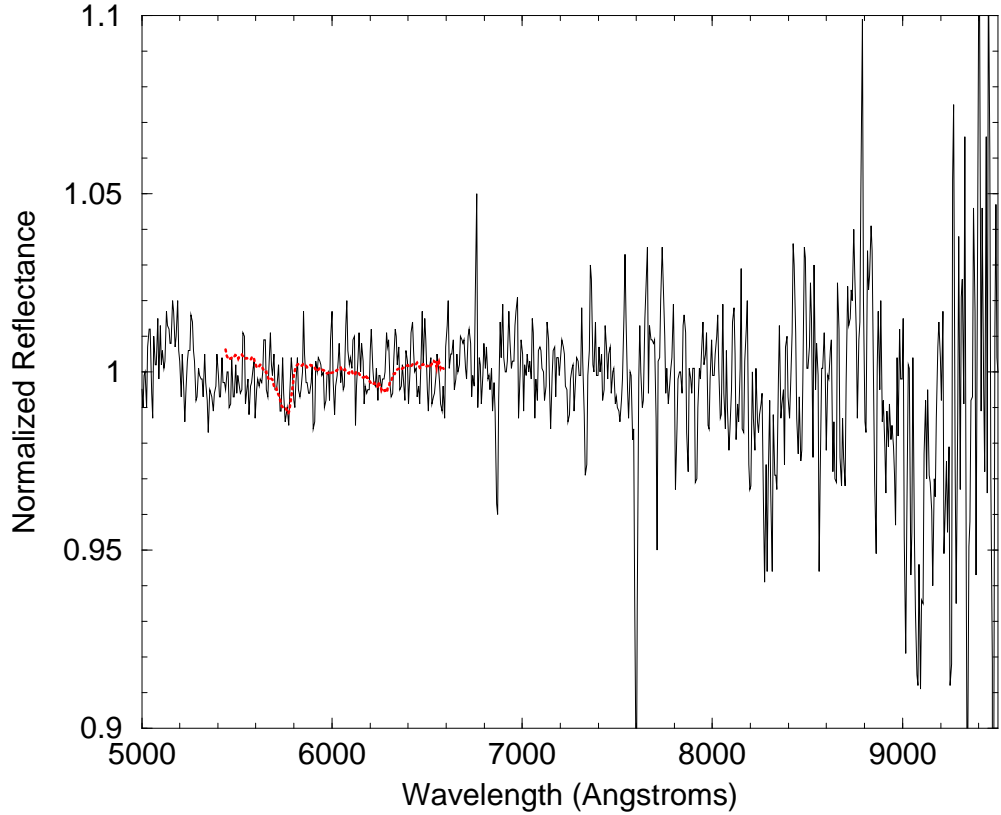}
\caption{Normalized reflectance spectrum of 2003 EL61 (black line) and
the spectral ratio of Ganymede to Callisto \citep{cal97} showing
O$_2$-ice bands at 5773 and 6275 \AA\ (red line).  There is a
tantalizing dip in the spectrum of 2003 EL61 at the position of the
5773 \AA\ band; however, a much higher continuum signal precision
is necessary to test for the presence of the weaker 6275 \AA\ band.}
\end{figure}

\clearpage

\begin{deluxetable}{lcc}
\tabletypesize{\scriptsize} \tablecaption{Laboratory Wavelengths and Frequencies of CH$_4$-Ice Absorption Maxima
\label{tbl-1}} \tablewidth{0pt} \tablehead{ \colhead{Transition} &
\colhead{Wavelength (\AA)} & \colhead{Wavenumber (cm$^{-1}$)}

}
\startdata
3$\nu_1$ $+$ 4$\nu_4$ & 7296 & 13706 \\
3$\nu_3$ $+$ 3$\nu_4$ & 7862 & 12719 \\
3$\nu_1$ $+$ 3$\nu4$  & 7993 & 12511 \\
4$\nu_3$              & 8415 & 11884 \\
$\nu_1$ $+$ 3$\nu_3$  & 8442 & 11846 \\
3$\nu_3$ $+$ 2$\nu_4$ & 8691 & 11506 \\
2$\nu_1$ $+$ $\nu_3$ $+$ 2$\nu_4$ & 8897 & 11240 \\
3$\nu_1$ $+$ 2$\nu_4$ & 8968 & 11151 \\
2$\nu_3$ $+$ 4$\nu_4$ & 9019 & 11088 \\
 \enddata
\tablenotetext{a}{Pure CH$_4$ Ice I at 30 K (Grundy et al. 2002)}
\end{deluxetable}


\clearpage

\begin{thebibliography}{}
\bibitem[Barkume et al.(2006)]{bar06} Barkume, K. M., Brown, M. E., and
Schaller, E. L. 2006, \apjl, 640, L87.
\bibitem[Brown et al.(2006)]{brn06} Brown, M. E., Barkume, K. M., Blake,
G. A., Schaller, E. L., Rabinowitz, D. L., Roe, H. G., \& Trujillo, C. A.
2006, \aj, submitted.
\bibitem[Brown et al.(2005)]{brn05} Brown, M. E., Trujillo, C. A., \&
Rabinowitz, D. L. 2005, \apj, 635, L97
\bibitem[Brown et al.(1999)]{brn99} Brown, R. H., Cruikshank, D. P., \&
Pendleton, Y. 1999, \apj, 519, L101
\bibitem[Calvin \& Spencer(1997)]{cal97} Calvin, W. M., \& Spencer,
J. R. 1997, Icarus, 130, 505.
\bibitem[Cruikshank et al.(1976)]{crk76} Cruikshank, D. P., Pilcher,
C. B., \& Morrison, D. 1976, Science, 194, 835
\bibitem[Cruikshank et al.(1993)]{crk93} Cruikshank, D. P., Roush, T. L., 
Owen, T. C., Geballe, T. R., de Bergh, C., Schmitt, B., Brown, R. H.
and Bartholomew, M. J. 1993, Science, 261, 742
\bibitem[Delitsky \& Lane(1997)]{del97} Delitsky, M. L., \& Lane,
A. L. 1997, J. Geophys. Res., 102, 16385
\bibitem[Doressoundiram et al.(2003)]{dor03} Doressoundiram, A.,
Tozzi, G. P., Barucci, M. A., Boehnhardt, H., Fornasier, S. \& Romon
J. 2003, \aj, 125, 2721
\bibitem[Elliot et al.(2005)]{ell05} Elliot, J. L., Kern, S. D.,
Clancy, K. B., Gulbis, A. A. S., Millis, R. L., Buie, M. W.,
Wasserman, L. H., Chiang, E. I., Jordan, A. B., Trilling, D. E., and
Meech, K. J. 2005, \aj, 129, 1117
\bibitem[Fink et al.(1980)]{fink80} Fink, U., Smith, B. A., Johnson,
J. R., Reitsema, H. J., Benner, D. C., \& Westphal, J. A. 1980,
Icarus, 44, 62
\bibitem[Fornasier et al.(2004a)]{for04a} Fornasier, S., Dotto, E.,
Barucci, M. A., \& Barbieri, C. 2004a, A\&A, 422, 43
\bibitem[Fornasier et al.(2004b)]{for04b} Fornasier, S.,
Doressoundiram, A., Tozzi, G. P., Barucci, M. A., Boehnhardt, H., de
Bergh, C., Delsanti, A., Davies, J., \& Dotto, E. 2004b,
A\&A, 421, 353
\bibitem[Grundy \& Buie(2001)]{gb01} Grundy, W. M., \& Buie,
M. W. 2001, Icarus, 153, 248
\bibitem[Grundy \& Fink(1993)]{gf93} Grundy, W. M., \& Fink, U. 1993, 
Pluto \& Charon Conference, Flagstaff, AZ, July 6-9, 1993
\bibitem[Grundy \& Fink(1996)]{gf96} Grundy, W. M. \& Fink, U. 1996,
Icarus, 124, 329.
\bibitem[Grundy et al.(2002)]{grun02} Grundy, W. M., Schmitt, B., and
Quirico, E. 2002, Icarus, 155, 486.
\bibitem[Hapke(1981)]{hap81} Hapke, B. 1981, J. Geophys. Res., 86,
4571.
\bibitem[Hapke(1993)]{hap93} Hapke, B. 1993, Combined Theory of
Reflectance and Emittance Spectroscopy (New York: Cambridge
Univ. Press)
\bibitem[Hardorp(1982)]{har82} Hardorp, J. 1982, A\&A, 105, 120
\bibitem[Jewitt \& Luu(2004)]{jl04} Jewitt, D. C., \& Luu, J. 2004,
Nature, 432, 731
\bibitem[Johnson(1990)]{jon90} Johnson, R. E. 1990, Energetic
Charged-Particle Interactions With Atmospheres and
Surfaces. (New York: Springer-Verlag)
\bibitem[Johnson \& Quickenden(1997)]{jon97} Johnson, R. E., \&
Quickenden, T. I. 1997, J. Geophys. Res., 102, 10985
\bibitem[Landau et al.(1962)]{lan62} Landau, A., Allin, E. J., \&
Welsh, H. L. 1962, Spectrochimica Acta, 18, 1.
\bibitem[Licandro et al.(2001)]{lic01} Licandro, J., Oliva, E., \& Di Martino, M. 2001, A\&A, 373, L29
\bibitem[Licandro et al.(2006)]{lic06a} Licandro, J., Pinilla-Alonso,
N., Pedani, M., Oliva, E., Tozzi, G. P., and Grundy, W. M. 2006,
A\&A, 445, L35.
\bibitem[Luu \& Jewitt(1996)]{lj96} Luu, J. \& Jewitt, D. 1996, \aj,
112, 2310
\bibitem[Massey et al.(1992)]{mas92} Massey, P., Valdes, F., \&
Barnes, J. 1992, in A User's Guide to Reducing Slit Spectra With IRAF
\bibitem[Quirico \& Schmitt(1997)]{qs97} Quirico, E., \& Schmitt, B. 1997,
Icarus, 127, 354
\bibitem[Rabinowitz et al.(2006)]{rab06} Rabinowitz, D. L., Barkume,
K., Brown, M. E., Roe, H., Schwartz, M., Tourtellotte, S., and Trujillo,
C. 2006, \apj, 639, 1238.
\end{thebibliography}
\end{document}